\documentclass[prb,aps,twocolumn]{revtex4}
\usepackage{epsf}
\usepackage{epsfig}
\usepackage{graphics}

\begin{document}

\title{A Magnetic Model for the Ordered Double Perovskites}

\author{Prabuddha Sanyal and Pinaki Majumdar}

\affiliation{Harish-Chandra  Research Institute,
 Chhatnag Road, Jhusi, Allahabad 211019, India}

\date{Dec 5,  2008}

\begin{abstract}
We construct an effective spin model from the coupled spin-fermion problem 
appropriate to double perovskites of the form A$_2$BB'O$_6$. The magnetic 
model that emerges is reminiscent of double exchange and we illustrate this 
`reduction' in detail for the case of perfect B-B' structural order, i.e, no 
antisite disorder. We estimate the effective exchange between the magnetic B 
ions in terms of the electronic parameters, study the `classical' magnetic 
model using Monte Carlo techniques, and compare this approach to a full 
numerical solution of the spin-fermion problem. The agreement is reasonable,
and promises a quick estimate of magnetic properties when coupled with  
{\it ab initio} electronic structure. The scheme generalises to the presence 
of antisite disorder.
\end{abstract}

\maketitle

\section{Introduction}

Double perovskite materials, of the form
A$_2$BB'O$_6$, have been of interest in recent years~\cite{DDreview,DDrev2} 
on account of their  magnetic, electronic and structural 
propeties. They promise large
magnetoresistance~\cite{GarciaLanda,Martinez,dds-prl-07},
potentially useful for switching applications. 
The half-metallic
character of some of the members also make them attractive
candidates for spintronic devices.

One of the species, B say, is typically magnetic, 
a transition metal like Fe, Co,
Ni, or Cr, while the B' species is generally 
non-magnetic, Mo, W, {\it etc}. 
The most studied 
member of this series is Sr$_{2}$FeMoO$_{6}$: it is 
a half-metallic ferromagnet (FM) 
at low temperature, and has a 
high $T_{c} \sim 410$K.
Sr$_{2}$FeWO$_{6}$, on the other hand, 
is an antiferromagnetic
(AFM) insulator! These limits illustrate the wide
range of physical properties in the double perovskites (DP). 
While the `endpoints' above  are relatively
easy to understand (ignoring disorder) there are
several effects where current understanding
is limited.

(i)~{\it Antisite disorder:}
Well annealed double perovskites tend to have an alternate
arrangement of B and B'
ions, but defects
called `antisite'  regions \cite{ps-pm-antisite}
appear
when two B or two B' atoms occur as neighbours.
These regions typically have
an AFM arrangement of the B spins and are insulating.
Their presence reduces the overall magnetization.
The electronic and magnetic properties in DP's are 
intimately related to the structural order.

(ii)~{\it Phase competition:}
Exploration of the
series~\cite{SugataBMoW,TokuraBMoW}
Sr$_{2}$FeMo$_{1-x}$W$_{x}$O$_{6}$ 
reveals a FM to AFM 
transition and an associated  metal-insulator transition 
(MIT) with increasing $x$.
In the regime of FM-AFM phase competition the
compounds show large magnetoresistance (MR).  

(iii)~{\it Magnetic B' sites:}
Recently, compounds where the B' site also has an intrinsic
magnetic moment have been investigated~\cite{Pickett}, 
and interesting compensation effects
have been observed. In particular, there are enigmatic compounds
 like Sr$_{2}$CrOsO$_{6}$ which are
insulating (semimetallic?), but at the same time ferromagnetic, 
with a very high $T_{c}$~\cite{Krockenberger}. In addition, there are spin-orbit 
effects~\cite{Vaithee} in some DP's complicating the
magnetic state. 

Apporaching issues (i)-(iii) above   
directly in a finite temperature
real space formulation is formidable. 
It requires tools that can predict magnetic properties of a 
double perovskite 
based on electronic parameters and the structural 
disorder.
This paper is a step towards that goal where we 
provide a semi-analytic 
scheme for accessing the magnetic ground state 
and $T_c$ scales
of a structurally ordered 
DP starting with a tight-binding spin-fermion model.
While our primary focus is the FM regime, we also
highlight issues of phase competition and antiferromagnetism
which are bound to be important when doping effects
are explored.

The paper is organised as follows.
The next section  describes the double perovskite model, 
following
which we summarise earlier work on this problem to place our 
work in context. We then 
outline the different methods used in this study. The
section after describes our results, primarily within a variational
scheme and an effective exchange calculation,
with Monte Carlo results
for benchmark. We then conclude, pointing out how our scheme
can be extended to the antisite disordered case.

\section{The double perovskite model}

The double perovskite structure of A$_2$BB'O$_6$
can be viewed as repetition of the perovskite units ABO$_3$
and AB'O$_3$. In the ideal ordered DP the B
and B' octahedra alternate in each direction.
In this paper
we  consider only the B ion to be magnetic.
The superexchange coupling between the B magnetic moments is
small in the ordered DP's.
The important physical ingredients in the problem are:
(i)~a large $S$ core spin at the B site,
(ii)~strong coupling on the B site between the core spin and
the itinerant electron, strongly  prefering  {\it one}
spin polarisation of the itinerant electron, and
(iii)~delocalisation of the itinerant electron on the B-O-B' network.

The Hamiltonian for the structurally (B-B') ordered double perovskites
is given by:
\begin{eqnarray}
H & =& \epsilon_{B}\sum_{i\in B}f_{i\sigma}^{\dagger}f_{i\sigma}+
\epsilon_{B'}\sum_{i\in B'}m_{i\sigma}^{\dagger}m_{i\sigma}
-\mu\sum_i (n_{f,i} + n_{m,i}) \cr
&&~  -t\sum_{<ij>\sigma}f_{i\sigma}^{\dagger}m_{j\sigma}
+ J\sum_{i\in A} {\bf S}_{i} \cdot
f_{i\alpha}^{\dagger}\vec{\sigma}_{\alpha\beta}f_{i\beta}
\label{fullhamSFMO}
\end{eqnarray}
The $f$'s refer to the magnetic B sites and the $m$ to the non magnetic
B', and the B-B' hopping 
$t_{BB'}=t$ is the principal hopping in the structurally
ordered DP's. We will discuss the impact of further neighbour hoppings
later in the text. We have retained only one orbital on the B and B'
site, our formulation readily generalises to a multiple orbital
situation.
The ${\bf S}_i$ are `classical' (large $S$)
 core spins at the B site, coupled
to the itinerant B electrons through a coupling $J \gg t$.
This implies that the conduction electron state at a B site is
slaved to the orientation of the corresponding B spin.
The difference between the ionic levels,
${\tilde \Delta} = \epsilon_B - \epsilon_{B'}$, defines the `bare' 
`charge transfer' energy.
At a later stage we will define the parameter $\Delta = (\epsilon_B
-JS/2) - \epsilon_{B'}$ as the `true' charge transfer energy.
$n_f$ is the B electron occupation number, while $n_m$ is the
B' electron occupation number.
We will assume $J/t \rightarrow
\infty$ (keeping $\Delta$ finite).
The parameter space of the problem is
defined by the electron filling, $n$, the ratio $\Delta/t$
and the temperature $T/t$. 
We have ignored 
Hubbard repulsion, 
B-B antiferromagnetic superexchange, and, to start with,
direct hopping between B'-B', or B-B.

\section{Earlier work}

Early work on the DP's was motivated by results on
Sr$_{2}$FeMoO$_{6}$, where 
electronic structure calculations indicate that 
Fe is in a $3d^{5}$ configuration (a half-filled state) 
while Mo is in
a $4d^{1}$ configuration. Following Hund's rule,
Fe is therefore in a high spin $S=5/2$ state.
Surprisingly, 
the normally nonmagnetic Mo picks up a moment of $1/2$ in the 
opposite direction, and reduces the moment per unit cell to  
$\sim 4 \mu_B$.
An explanation for the induced moment on the non 
magnetic B'  species was provided \cite{DDPRL}
by Sarma {\it et al}, in terms of a `level repulsion'
between the
Fe and Mo levels. Such a scenario implies a 
substantial degree of hybridization
between the Fe and Mo orbitals, and  
assumes that the itinerant Mo electron
hops through the Fe sublattice. 

Using this idea, a double-exchange (DE) 
like 2-sublattice Kondo
lattice model was proposed for the DP's~\cite{Millis}, 
and solved within dynamical mean field theory by Chattopadhyay 
and Millis~\cite{Millis},
assuming a `ferrimagnetic' \cite{ferri} state. 
They obtained a $n-T$ phase
diagram for different values of  
$\epsilon_{B'}-\epsilon_{B}$ and $J$ and 
observed that the $T_{c}$ goes to zero at 
large filling, indicating the presence of some competing
non ferromagnetic  state. 

A similar result for $T_{c}(n) $
was obtained by Carvajal {\it et al}~\cite{Avignon} 
using  another two sublattice model,
and Ising spins.
Here the hopping of an electron 
with spin $\sigma$ from a B'  site to 
a neighbouring B site is $t$ if $\sigma$ is 
antiparallel to the local spin $\mu_{i}$ on that
site, while it is zero if they are parallel. 
The authors considered only ferrimagnetic arrangements. 

Alonso 
{\it et al}~\cite{FGuinea} considered a variant of Millis' model 
with the coupling $J \rightarrow \infty$,
but with a larger number of ordering possibilities.
They also took into account possible 
antisite defects, including  a B-B hopping 
and superexchange which are only active when two B atoms become 
nearest neighbours.
They considered four possible phases: (1)~paramagnetic, 
(2)~ferrimagnetic, (3)~an AFM phase,
 where the B spins in neighbouring (1,1,1) planes are 
antiparallel, and (4)~another 
ferrimagnetic phase where the B spins are 
aligned ferromagnetically if the B are in the correct 
positions, and antiferromagnetically if the B ions occupy B' sites 
due to antisite defects. 
Among other results they 
found that even in the B-B' ordered case (where  
superexchange is not operative) 
the AFM phase is preferred to the FM
at high band filling.
 
All these studies, except the paper by Alonso {\it et al.},
concentrate on the ferromagnetic \cite{ferri}
phase. They observe the 
decrease of $T_{c}$ at large filling
but do not explore competing phases.
Secondly, while the DMFT approaches provide a 
semianalytic treatment of the $T_{c}$ scales, in specific parts 
(in this case ferromagnetic)
of the phase diagram, an estimate of the effective exchange between
the B moments is not available. 
The {\it ab initio} approaches have attempted such
an estimate by force fitting a `Heisenberg model'. 
Unfortunately, the
magnetic states that emerge from the DP model, and the effective exchange
that stabilises these phases arise from subtle electron delocalisation
physics not captured by such methods. 
We also do not know of any work that allows an economnical and
systematic exploration of the parameter space, $n$, $\Delta$, $J$, of
the DP model.
The present paper aims to overcome these shortcomings.

\section{Methods}

The first estimate of magnetic interactions in any material is
provided by {\it ab initio} calculations. This is typically done by calculating
the difference in ground state energy of the compound in spin polarized
and spin-unpolarized configurations; or in different magnetic ground states
 corresponding to different values of the spin density wave vector~\cite{DDPRL}. 
Such a calculation involves
all the relevant orbitals and their hybridization and provides
a rough material specific estimate.
However, for  complex
antiferromagnetic ground states one has to guess such
configuration beforehand, or take a cue from experiments. 
There is  no {\it a-priori} prescription for finding them.

Model Hamiltonian based calculations, on the other hand, have the obvious 
limitation
that model parameters have to be inferred from elsewhere, typically 
{\it ab initio} studies~\cite{OKAndersen, Tanusri}.  The advantage,
however, lies in the simplicity of the resulting model, and our 
ability to create a qualitative understanding using the tools of 
statistical mechanics. 
The Hamiltonian appropriate to double perovskites 
can be studied using the following tools: (i)~a combination of 
exact diagonalization  and Monte Carlo (ED-MC), (ii)~variational
calculation (VC) based on some family of periodic spin configurations,
and (iii)~mapping to an effective classical spin model.

The ED-MC approach has the advantage of accessing the
magnetic structure without 
bias. However, due to large computational
cost, it is severely size-limited, limiting the class of
magnetic structures which can be probed. A `travelling cluster'
(TCA) variant \cite{tca-ref}
 of ED-MC allows use of somewhat larger system size.
Variational calculations assuming a periodic spin background can be used
for very large system size (since there is no bulk diagonalisation
needed) but are restricted by the choice of the variational family.
While we will use both (i) and (ii) above, our principal tool
will be (iii), where we map on the spin-fermion problem to
an effective spin only model, with exchange calculated from the
fermions \cite{pm-scr}. 
We describe (i)-(iii) in more detail below.

\subsection{Monte Carlo}

One can
solve the DP model on a finite lattice  by direct numerical methods,
allowing for an `exact' benchmark for approximate solutions.
ED-MC is such a technique. Here, the coupled
spin-fermion problem is solved
by updating the classical spins using a Monte Carlo, 
diagonalizing the fermion system at each
step of the MC to infer the energy cost of the move. 
The method is numerically
expensive and can only be used on  
small system sizes, $\sim 8 \times 8$. Substantially bigger sizes,
$\sim 24 \times 24$, can be accessed using the TCA.

For the MC implementation 
the Hamiltonian of Eq~\ref{fullhamSFMO} has to be cast
into
form appropriate for  $J\rightarrow\infty$. This is done by 
performing a rotation to the local ${\bf S}_i$  axis at each
B  site, and retaining only the electron state 
oriented antiparallel to ${\bf S}_i$ at that
 site. This gives the following
Hamiltonian, with `spinless' B conduction electrons
and B' electrons having both spin states.
\begin{eqnarray}
H &=& t\sum_{<ij>}
\{ ( sin({{\theta_i} \over 2})f_{i}^{\dagger}m_{j\uparrow} 
 -
e^{i\phi_i}cos({{\theta_i} \over 2})f^{\dagger}_{i}m_{j\downarrow}) +h.c.\}
 \cr
&&~~~~~~~~~
+\epsilon_{B}\sum_{i}f_{i}^{\dagger}f_{i}+
\epsilon_{B'}\sum_{i\sigma}m_{i\sigma}^{\dagger}m_{i\sigma}
\label{Jinfinityham}
\end{eqnarray}

There is no longer any `infinite' coupling in the model,
and the number of degrees of freedom has been reduced to
one per B site (and 2 per B'), so the Hilbert space is
a little smaller. $m_{j\downarrow}$ and $m_{j\uparrow}$
hop to different conduction electron projections at the
neighbouring B site(s) so the effective hopping picks up
a $\theta_i, \phi_i$ dependent modulation.
We will use this form of the DP model for the Monte Carlo.

\subsection{Variational ground state}

A more analytical method  used before in the double 
exchange context is to write down a family of spin 
configurations $\{ {\bf S} \}_{\alpha}$, denoted  $S_{\alpha}$
for simplicity, 
and calculate the electronic energy in that background. Since the
$S_{\alpha}$ are usually periodic this is effectively a `band structure'
calculation. For a specified chemical potential one can calculate
the electronic energy 
${\cal E}(\mu, S_{\alpha})$. The configuration 
$S_{min}(\mu)$ that minimises ${\cal E}$ is the variational
ground state. 
Needless to say,  the `minimum' is only as good as 
the starting set, and in general non periodic $S_{\alpha}$
cannot be handled.
Nevertheless, used in combination with MC results 
it can be a valuable tool.

From the MC we will discover that in the structurally ordered case
the DP model has simple periodic ground states, with
windows of phase separation in between. This will allow
us to use the variational scheme, with only a few 
configurations, to map out the $T=0$ phase diagram
accurately.

\subsection{Effective exchange}

The complications with spin-fermion MC, and the
limitations of VC could be avoided if one had an explicit spin-spin
interaction model deduced from the starting DP model.
Formally such a scheme can be written down, and some
progress made through approximation.
Let us illustrate this `self consistent renormalisation' 
(SCR)  principle \cite{pm-scr}  in the simpler context of
double exchange before moving to the double perovskites.

\subsubsection{Illustrative case:  double exchange model}

Consider the following model:
\begin{equation}
H= \sum_{ij\sigma}t_{ij} 
c_{i\sigma}^{\dagger}c_{j \sigma} - J\sum_{i,\alpha\beta}
{\bf S}_i \cdot c_{i\alpha}^{\dagger}\vec{\sigma}_{\alpha\beta}c_{i\beta}
\end{equation}
Let us try to construct
an approximate classical spin model in the limit $J \rightarrow\infty$.
The classical model is defined by the equivalence:
\begin{equation}
\int {\cal D}  {\bf S}_i e^{ -\beta H_{eff} \{ {\bf S} \}}
=
\int {\cal D}  {\bf S}_i Tr e^{-\beta H}
\end{equation}
where the trace is over the fermion degrees of freedom. 
The trace, in general, is impossible to compute analytically
since it involves the spectrum of fermions moving in an
{\it arbitrary} spin background $\{ {\bf S} \}$. 
Nevertheless, some headway can
be made once the Hamiltonian is written in a more suggestive 
rotated and projected basis as \cite{pm-scr}:
\begin{equation}
H= \sum_{ij}f_{ij} 
t_{ij} (e^{i\Phi_{ij}}\gamma_{i}^{\dagger}\gamma_{j}
~+~h.c)
\end{equation}
where 
$f_{ij}=\sqrt{\frac{1+ {\bf S}_{i}\cdot {\bf S}_{j}}{2}}$, $\Phi_{ij}$
is a phase factor depending on ${\bf S}_i$ and ${\bf S}_j$, and the
$\gamma$ are `spinless' fermion operators. 
This suggests the approximation:
\begin{eqnarray}
H_{eff} \{ {\bf S} \} &\approx & -\sum_{ij} D_{ij} 
\sqrt{\frac{1+ {\bf S}_i.{\bf S}_j }{2}} \cr
D_{ij} &=& -t_{ij} 
\langle \langle e^{i\Phi_{ij}}\gamma^{\dagger}_{i}\gamma_{j} + h.c 
\rangle \rangle 
\label{eff_spin_model_DE}
\end{eqnarray}
The angular brackets indicate first a quantum average 
(for fixed $ \{ {\bf S} \}$) and then thermal
average over $e^{-\beta {H_{eff}\{ {\bf S} \}}}$.

Another way to obtain the same result, which 
generalises to the DP problem, is to write the
action for $H$ in a spin background $\{ {\bf S} \}$:
\begin{equation}
{\mathcal A} \{ {\bf S} \} =\beta\sum_{n,i,j}
\left[i\omega_{n}\delta_{ij} - t_{ij}
 f_{ij}e^{i\Phi_{ij}}\right]
\gamma_{in}^{\dagger} \gamma_{jn}
\end{equation}
and the internal energy 
$U \{ {\bf S} \}=\frac{\partial lnZ\{ {\bf S} \}}
{\partial\beta}=- \langle 
\frac{\partial {\mathcal A}}{\partial\beta}\rangle $;
\begin{eqnarray}
U\{ {\bf S} \} &=& \sum_{ij}f_{ij} t_{ij} e^{i\Phi_{ij}}\sum_{n}
\langle \gamma^{\dagger}_{in}\gamma_{jn} \rangle  \cr
&=& \sum_{ij}f_{ij} t_{ij} ( e^{i\Phi_{ij}}
\langle \gamma^{\dagger}_{i}\gamma_{j} \rangle  + h.c)
\end{eqnarray}
which is simply the quantum average of the spin-fermion Hamiltonian
for a fixed $\{ {\bf S} \}$. 
As before we 
can convert this to an approximate spin Hamiltonian by
thermally averaging the quantity within the round brackets.

The effective exchange depends on $T$, but in the `clean'
problem it does not depend on the `bond' $ij$. 
Since the DE model always has a ferromagnetic
ground state, the low
$T$ exchange can be calculated from the fermionic
average in the {\it fully polarised} state, and is
simply: 
$$
D  \propto 
\sum_{k}\epsilon_{k} \langle n_{k} \rangle  =
\sum_{k}\epsilon_{k}n_{F}(\epsilon_{k}) 
$$
It was observed \cite{pm-scr} that even at finite
temperature the {\it self consistent average} in the ferromagnetic phase
remains close to the $T=0$ value till very near $T_{c}$. 
The $T=0$ kinetic energy therefore provides a reasonable 
estimate of effective ferromagnetic exchange, and so the 
$T_{c}$.
The overall scale factor between the $T_{c}$ and the exchange 
can be determined from a Monte Carlo calculation.

\subsubsection{Effective exchange in the double perovskites}

Unlike the DE model we cannot write an effective spin only 
Hamiltonian for
the double perovskites purely by 
inspection since the electron motion also
involves the B' sites. We use the action formulation instead.
Integrating out the B' electrons 
 we get an action entirely in terms of the
B degrees of freedom:
\begin{eqnarray}
{\mathcal A}\{ {\bf S} \} &=& 
\beta\sum_{n}
(\sum_{k\sigma}
f_{kn\sigma}^{\dagger}G_{ff0}^{-1}(k,i\omega_{n})f_{kn\sigma} \cr
&&~~~~~~~~~~~~ -J\sum_{i} {\bf S}_{i} \cdot f_{in\mu }^{\dagger}
\vec{\sigma}_{\mu \nu } f_{in\nu })
\label{B_action}
\end{eqnarray}
where $G_{ff0}(k,i\omega_{n})$ is the $J=0$  Greens function
involving B sites only (the $n$ represent Matsubara frequencies):
\begin{equation}
G_{ff0}^{-1}=i\omega_{n}-(\epsilon_{B}- \mu)
- \frac{\epsilon_{k}^{2}}{i\omega_{n}-(\epsilon_{B'} -\mu)}
\end{equation}
If we choose $\epsilon_{B}=0$,~$\epsilon_{B'}=\Delta<0$, this becomes:
\begin{equation}
G_{ff0}^{-1}=i\omega_{n} + \mu - 
\frac{\epsilon_{k}^{2}}{i\omega_{n} + \mu -\Delta}
\end{equation}
where  $\epsilon_{k}=2t \sum_{i=1}^{d}cosk_{i}a$.
The poles of this Greens function give the band dispersion 
at $J=0$:
\begin{equation}
E_k^{\pm}=\frac{\Delta\pm\sqrt{\Delta^{2}+4\epsilon_{k}^{2}}}{2}
- \mu
\label{disp_eqn_tFM}
\end{equation}
In the limit $\Delta \gg t$, i.e., the limit of weak charge transfer, 
there are two bands centred roughly on $0$ and $\Delta$. 
For $\Delta=0$, there are two bands
$\pm|\epsilon_{k}|$ symmetrically placed about 0.

While the first term in the action involving this bare Greens function
conserves spin and momentum, the
second term is local in real space and typically involves spin-flip.
To proceed, let us Fourier transform $G_{ff0}^{-1}(k,\omega)$ 
and write the action in real space.
$\epsilon_k^2$ generates `hoppings' (in the full B-B' lattice) 
connecting sites that can either be next nearest
neighbours (2N), next-to next nearest neighbours (3N), 
or the same site. 
%({\bf show a lattice indicating these neighbours}).
In real space the action assumes the form:
\begin{eqnarray}
{\mathcal A} \{ {\bf S} \} &=&
\beta\sum_{n} (\sum_{ij\sigma}f_{in\sigma}^{\dagger}
G_{ff0}^{-1}(\vec{r}_{i}-\vec{r}_{j},i\omega_{n})f_{jn\sigma}  \cr
&&~~~~~~~~~~~~ - J\sum_{i} {\bf S}_i .  
f_{in\alpha}^{\dagger}\vec{\sigma}_{\alpha\beta}f_{in\beta})
\label{realsp_action}
\end{eqnarray}
Now, an unitary transformation is performed in spin space so that
the second term in the action becomes diagonal:
$
\gamma_{i n \mu  }=\sum_{\alpha}A_{\mu\alpha}^{i}f_{i n \alpha}.
$
The action becomes:
\begin{eqnarray}
{\mathcal A} \{ {\bf S} \} &=&
\beta\sum_{n}(
\sum_{ij\mu,\nu\sigma}
g_{\mu \nu}^{ij} 
\gamma_{i\mu n}^{\dagger}G_{ff0}^{-1}(\vec{r}_{i}-\vec{r}_{j},i\omega_{n})
\gamma_{j\nu n} \cr
&&~~~~~~~~~ -{{JS} \over 2}\sum_{i}
(\gamma_{iun}^{\dagger}\gamma_{iun}-\gamma_{iln}^{\dagger}\gamma_{iln})
)  
\label{gij_action}
\end{eqnarray}
where
$
g^{ij}_{\mu\nu}=\sum_{\sigma}A^{i}_{\mu\sigma}A^{j^{\dagger}}_{\sigma\nu}
$.

%--------------------------------------------------------------------
\begin{figure*}
\includegraphics[width=4.5cm,height=4.5cm,angle=-90,clip=true]{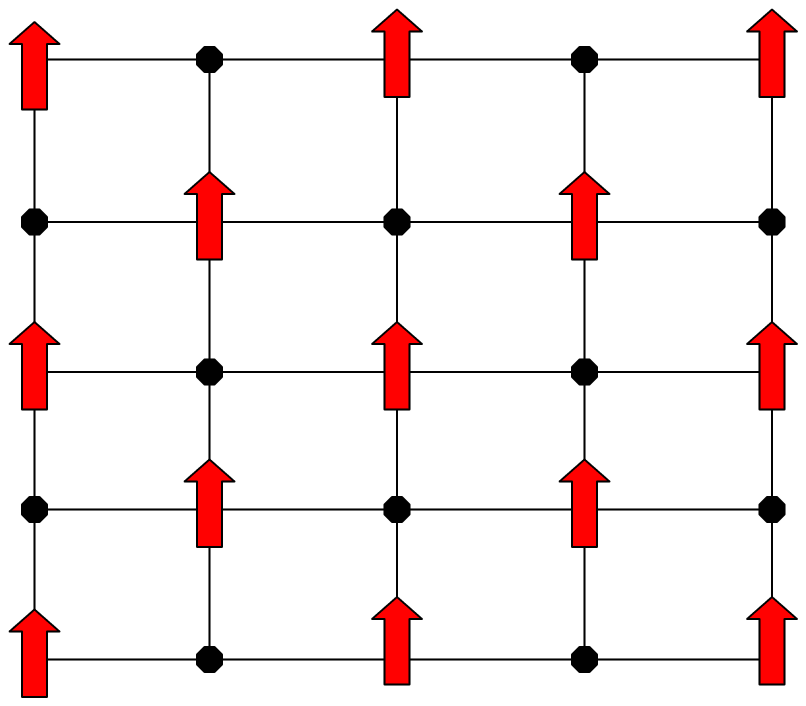}
\hspace{.9cm}
\includegraphics[width=4.5cm,height=4.5cm,angle=-90,clip=true]{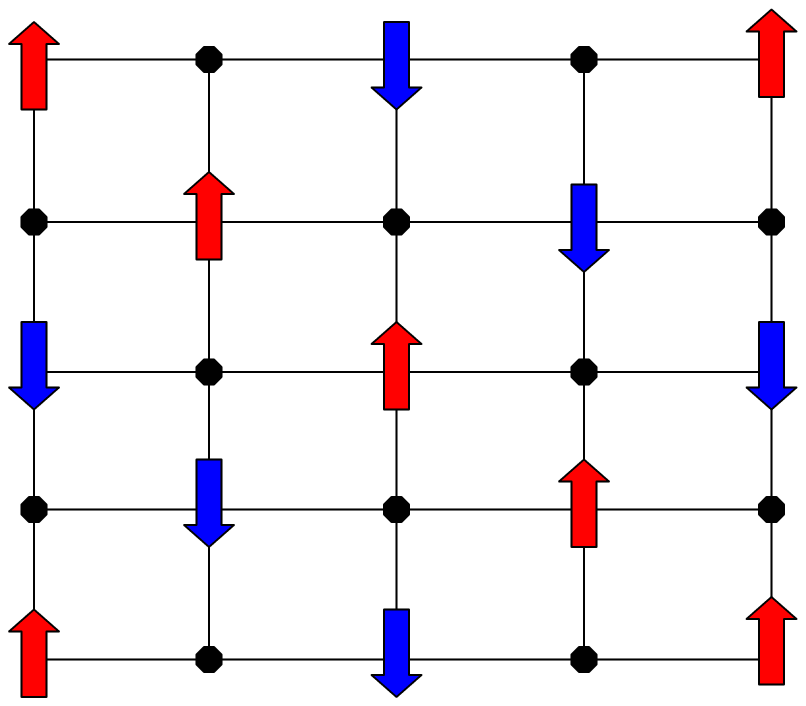}
\hspace{.9cm}
\includegraphics[width=4.5cm,height=4.5cm,angle=-90,clip=true]{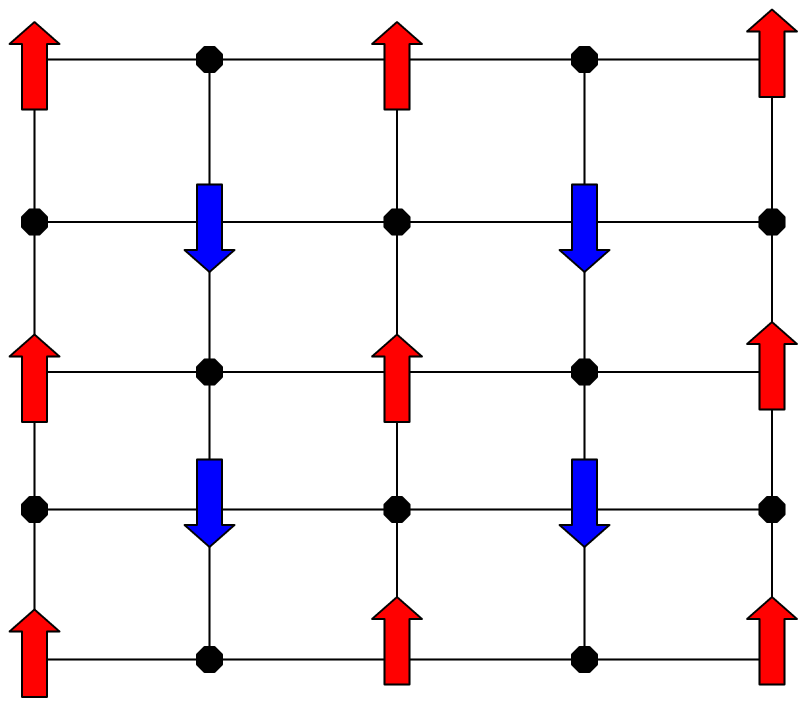}
\caption{Colour online: The three magnetic phases in the 2D model.
Left: ferromagnetic (FM), center: 
antiferromagnet (AFM1), right:  antiferromagnet 
(AFM2). These occur with increasing
electron density. The moments are on the B sites,
we have not shown the induced moments on the B' sites.}
\label{fig: phases}
\end{figure*}
%--------------------------------------------------------------------

At large $J$ one projects out the $\gamma_{iu}$ states,
retaining only the terms involving the
index $l$. 
Thereafter, we drop this index, 
redefine the B level as $\epsilon_B 
\rightarrow  \epsilon_B - {{JS} \over 2}$, and
obtain an effective spinless fermion
model similar to the case of double exchange.
\begin{eqnarray}
A\{ {\bf S} \} &=&
\beta\sum_{ij}g_{ij}\overline{\gamma}_{in}
G_{ff0}^{-1}(\vec{r}_{i}-\vec{r}_{j},i\omega_{n})\gamma_{jn}  \cr
&=& \beta\sum_{ij}g_{ij}
((i\omega_n + \mu)  \delta_{ij}
-\frac{h_{ij}}{i\omega_{n} + \mu -\Delta})
\overline{\gamma}_{in}\gamma_{jn}
\nonumber
\end{eqnarray}
where $g_{ij}=\sqrt{\frac{1+ {\bf S}_{i} \cdot {\bf S}_{j}}{2}}
e^{i\Phi_{ij}}$ as before.  $ h_{ij}$ is the
Fourier transform of $\epsilon_k^2$ and connects
sites on the B sublattice. It involves
a NN term (${\hat x} + {\hat y}$ in the full B-B' lattice) and
a third neighbour term ($2 {\hat x}$ etc in the B-B' lattice).

It is important we appreciate the various terms in the expression
for $A\{ {\bf S} \}$ above. The `kernel' 
$G_{ff0}^{-1}(\vec{r}_{i}-\vec{r}_{j},i\omega_{n})$ 
is specified by the $J=0$ bandstructure of the B-B' problem, 
explicit information about the spin variables is encoded in $g_{ij}$,
and the fermions are defined {\it in the background $\{ {\bf S} \}$ }.

We define  $i\nu_{n}=i(2n+1)\pi$, so 
$$
A\{ {\bf S} \} = \sum_{in}
(i\nu_{n} + \beta \mu) \overline{\gamma}_{in}\gamma_{in}-
\sum_{ijn} g_{ij} \frac{\beta^{2} h_{ij}}
{i\nu_{n} + \beta \mu - \Delta}\overline{\gamma}_{in}\gamma_{jn}
$$
The internal energy can be calculated as
$U \{ {\bf S} \} = -\frac{\partial lnZ}{\partial\beta}=
- \langle \frac{\partial A\{ {\bf S} \}}{\partial\beta}\rangle $.
Simplifying the resulting expression and 
using the same principle as in DE we can write 
an explicit (but approximate) model
purely in terms of core spins:
\begin{eqnarray}
H_{eff} \{ {\bf S} \} &=&
\sum_{ij} D_{ij} \sqrt{ {1+ {\bf S}_i.{\bf S}_j \over {2}}}  \cr
D_{ij}& =&  h_{ij} {{1} \over {\beta}} \sum_n B(i\omega_n) 
\langle e^{i\Phi_{ij}} 
\langle \overline{\gamma}_{in}\gamma_{jn} \rangle  + h.c
\rangle  \cr
B(i\omega_n) & = & 
{ { (2i\omega_{n} + 2 \mu -\Delta)} \over {(i\omega_{n} + \mu -\Delta)^{2}}}
\label{spinmodel}
\end{eqnarray}
The effective exchange $D_{ij}$ can be determined 
at any temperature by the SCR
method. 
The couplings take two values, $D_1$ for NN B-B exchange, and $D_2$ for
second neighbour B-B exchange.
The low $T$ exchange can be estimated by evaluating the fermionic average 
in the perfectly spin-ordered state at $T=0$ (after checking that the
ground state generated by the exchange is self-consistently ferromagnetic).
The evaluation of the Matsubara sums, {\it etc},  is
discussed in Appendix~A.

\section{Results}

\subsection{The magnetic ground state}

Both ED-MC done on $8 \times 8$, and TCA done on $16\times 16$ exhibit the
presence of three phases:
 namely ferromagnetic (FM, first panel in Fig.1), a `line like'
antiferromagnetic phase (AFM1, middle panel) 
and the more conventional antiferromagnet (AFM2) in
the last panel. If we define the ordering wave-vector on
B sublattice using  axes along the
diagonals, the FM phase has  
order at ${\bf Q} = \{0, 0\}$, AFM1 has order at
${\bf Q} = \{0, \pi \}$, and AFM2 has order at
${\bf Q} = \{\pi, \pi \}$.
We could of course define the wave-vectors on the full
B-B' lattice and use the usual $x$ and $y$ axes, but
for the structurally ordered case the earlier convention 
is simpler.

\begin{figure}[b]
\includegraphics[width=6cm,height=7.5cm,angle=0,clip=true]{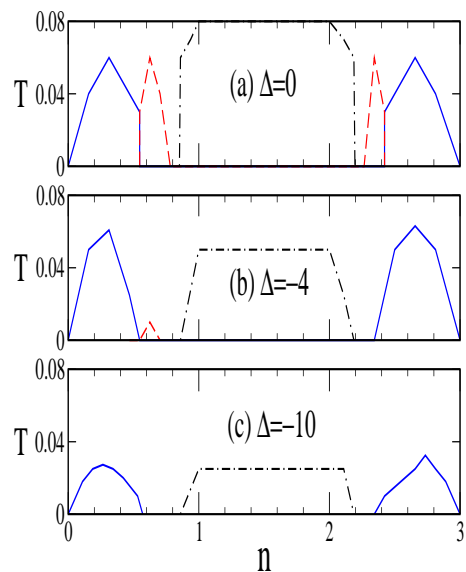}
\caption{Colour online:
$n-T$ phase diagram based on TCA. 
(a)~$\Delta=0$, (b)~$\Delta=-4$, and (c)~$\Delta=-10$.
The solid (blue) lines are ferro, dashed (red) lines are AFM1 and 
the dash-dot (black) lines are AFM2. The system size is $16 \times 16$.
}
\label{fig: TCA_phase_diag}
\end{figure}

Fig~\ref{fig: TCA_phase_diag} shows the $n-T$
phase diagram for three values of $\Delta$.
With increasing $n$ the phases occur in the
sequence FM, AFM1, AFM2, AFM1 and FM again.
The sequence as
well as the rough filling windows are
similar for all three $\Delta$ values (the
VC, which is free of size effects, will demonstrate this
more clearly).
Fig~\ref{fig: TCA_phase_diag}(b) shows that
for intermediate level difference, $\Delta=4$,
the $T_{c}$ for the ferromagnetic phases actually
increase a little bit.
The $T_{c}$ of the
antiferromagnetic phases, however, decrease. Moreover, the $\{ 0, \pi\}$ phase is
unobservable on the high filling side, while its Neel temperature, $T_{N}$,
 is quite small even on the
low filling side.
Eventually, for large enough $\Delta$, the
$T_{c}$ of even the ferromagnetic phases decrease,
as seen in Fig~\ref{fig: TCA_phase_diag}(c).
The AFM1 phase is unobservable on even
the low filling side, possibly due to very small $T_{N}$.

While AFM phases driven by B-B superexchange have been studied in
the DP's, AFM phases {\it driven by electron delocalisation} have not
seen much discussion. Their occurence, however, is not surprising.
If we were to `test out' the feasibility of various magnetic ground
states we could restrict ourself to a few simple collinear phases
to start with.
The FM, AFM1, etc, are such examples. Let us index them by
some index $\alpha$. As described before, which of these occur
at a chemical potential $\mu$ can be simply checked by
calculating the energy 
${\cal E}_{\alpha}(\mu) = \int_{-\infty}^{\mu} d\epsilon
N_{\alpha}(\epsilon) \epsilon$, where $N_{\alpha}(\epsilon)$
is the electronic density of states in the spin background
$\alpha$. 
The phase appropriate to a particular $\mu$ would be
the one with lowest energy.
Even without a calculation it is obvious that the FM state will 
have the largest bandwidth, and would be preferred at low
$n$. The AFM phases have narrower bands, but larger density of
states (since the overall DOS is normalised), 
and with growing $\mu$ they
become viable. In what follows we quantify this carefully.

%-------------------------------------------------------------------
\begin{figure}[b]
\includegraphics[width=6cm,height=5cm,angle=0,clip=true]{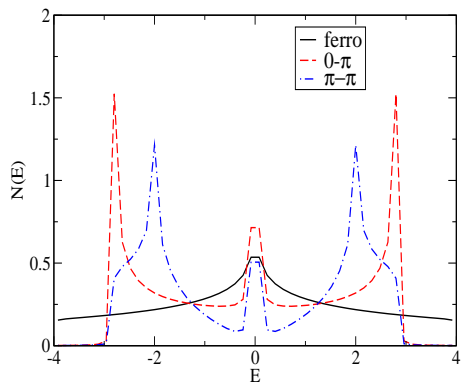}
\caption{Colour online: 
Electronic density of states
for the three variational states, with
${\bf Q}=  \{0, 0\},~\{ 0, \pi \}$ and $\{\pi, \pi\}$
at~$\Delta=0$}
\label{DOS_ferro_0pi_pipi}
\end{figure}

%-------------------------------------------------------------------
\begin{figure}[h]
\includegraphics[width=7cm,height=7cm,angle=0,clip=true]{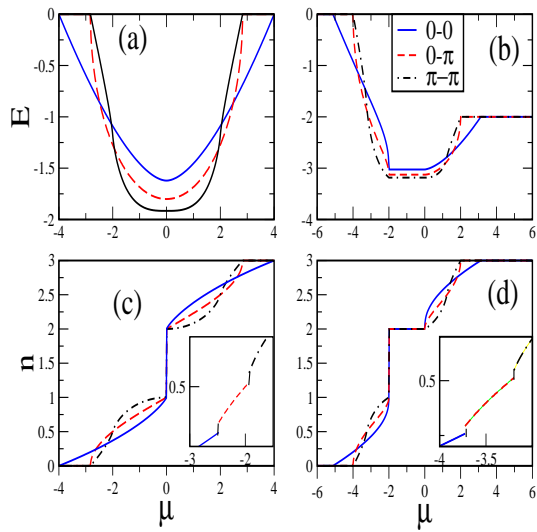}
\caption{Colour online: 
Electronic energy (top) and
filling $n$ (bottom) versus chemical potential for
the three phases, $\Delta=0$ (left), $\Delta=-2$ (right)}
\label{fig: E_ferro_0pi_pipi}
\end{figure}
%-------------------------------------------------------------------

One can obtain analytic expressions for the
dispersions in the $\{0, \pi\}$  and $\{\pi, \pi \}$ phases,
which are given below. In the $\{0, \pi\}$ phase,
in our $J\rightarrow \infty$ limit, the
structure decomposes into electronically decoupled
ferromagnetic zigzag chains aligned antiferromagnetically
with respect to each other, see Fig.1 middle panel.
Their dispersion is 1D-like, given by:
\begin{equation}
\epsilon_{\bf k}=\frac{\Delta\pm\sqrt{\Delta^{2}+16+16cos(k_{x}-k_{y})}}{2}
\end{equation}
In the limit $\Delta\rightarrow 0$, their 1D like nature is clearly visible:
\begin{equation}
\epsilon_{\bf k}=2\sqrt{2}cos({{k_{x}-k_{y}} \over 2})
\end{equation}

The $\{\pi, \pi \}$ phase, on the other hand,
decouples into two planar lattices
where the B spins are arranged ferromagnetically,
while these lattices are themselves aligned
antiferromagnetically with respect to each other.
The dispersion is given by:
\begin{equation}
\epsilon_{\bf k}=\frac{\Delta\pm\sqrt{\Delta^{2}+16t^{2}(cos^{2}k_x+cos^{2}k_y)}}{2}
\end{equation}
which reduces to $\pm2t\sqrt{cos^{2}k_x+cos^{2}k_y}$
when $\Delta=0$. It is interesting to
 note that the bandwidths of both $\{0, \pi\}$ and $\{\pi, \pi \}$
phases are identical, although the detailed DOS are
different.
The DOS for the three phases for
$\Delta=0$ are shown in Fig~\ref{DOS_ferro_0pi_pipi}.
We can understand the occurence of the various phases 
by integrating the DOS and comparing the energies
at a fixed $\mu$.
The results are shown in 
Fig~\ref{fig: E_ferro_0pi_pipi}(a) for $\Delta=0$ and Fig~\ref{fig: E_ferro_0pi_pipi}(b) for $\Delta=-2$.

As expected, at low filling the 
energy of the FM phase is the lowest,
while for intermediate filling that of the 
AFM1 phase is lower than the FM phase.
At still higher fillings, the energy of 
the AFM2 phase is the lowest. This is repeated
symmetrically on the other side of 
$\mu=0$ for $\Delta=0$. 
The density discontinuity corresponding to each transition 
can also be found from the corresponding
$\mu-n$ curves. 
For finite $\Delta$, the AFM1 
phase becomes narrower,
especially on the high filling side. These simple variational
results are corroborated by the phase diagram obtained from 
the TCA calculation. 

\begin{figure}[t]
\includegraphics[width=7cm,height=4.5cm,angle=0,clip=true]{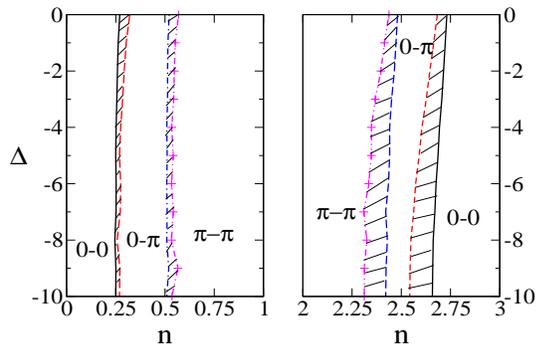}
\caption{Colour online:
$n-\Delta$ phase diagram at $T=0$ 
from the variational calculation.
The hashed regions indicate windows of phase separation between the
adjoining phases.
}
\label{fig: phasediag_DeltaN}
\end{figure}

The $n-\Delta$ phase diagram  at $T=0$ 
is shown in Fig~\ref{fig: phasediag_DeltaN}.
There are windows of phase separation (PS) where homogeneous
electronic/magnetic states are not allowed.
These regions correspond to the jumps in the $n-\mu$ curve.
The AFM1 
phase becomes unstable on the high filling side for 
large $\Delta$, which manifests itself through  
a merging of the phase boundaries.

\subsection{Spin model and effective  exchange}

Our effective spin model is:
$$
H_{eff} \{ {\bf S} \} =
\sum_{ij} D_{ij} \sqrt{ {1+ {\bf S}_i.{\bf S}_j \over {2}}} 
$$

\begin{figure}[t]
\includegraphics[width=5.5cm,height=6.5cm,angle=0,clip=true]{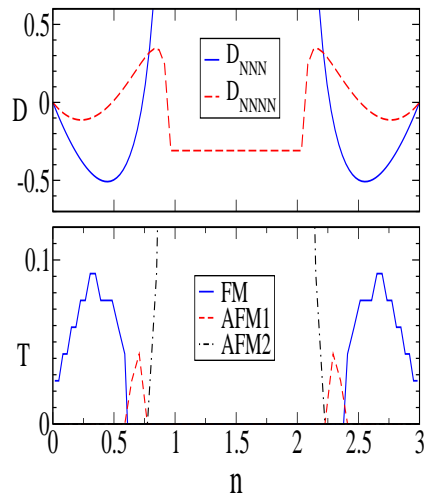}
\caption{Colour online: 
$D_{ij}$ for NN and NNN, $\Delta=0$ (top) and 
$n-T$ phase diagram obtained from
a Monte Carlo on the spin model using this exchange (bottom).}
\label{fig: Dij_Delta0_2}
\end{figure}

Using the exchange $D_{ij}$ calculated from 
Eq~\ref{spinmodel} using a fully  ferromagnetic reference state, 
one can plot
the nearest neighbour exchange $D_{NN}$ and the 
next nearest neighbour $D_{NNN}$ as a 
function of filling $n$. 
The results are shown in the top panel of Fig~\ref{fig: Dij_Delta0_2}(a) 
for $\Delta=0$. 
One 
finds that both the exchanges change sign
as a function of filling. At low filling and very high filling, 
both are negative, indicating an overall
ferromagnetic coupling. However, for intermediate values of 
filling, both the exchanges become
positive, giving an effective antiferromagnetic coupling. 
In between, there is a small
region where one of them is positive and the other negative. 

Since the calculation was started
using a purely ferromagnetic spin background, such changes 
in sign of the calculated exchange 
indicate an instability of the ferromagnetic phase at these 
fillings. Where the `exchange' $D_{NN} + D_{NNN} > 0$ the ground state
will no longer be FM, the result for $D_{ij}$ 
is not self-consistent, and the quantitative values not
trustworthy.
We will confine ourself to the window where the ground state is
self consistently ferromagnetic.

The exchange for the antiferromagnetic states, and the Neel temperature,
should be calculated in appropriate spin backgrounds, 
i.e., $\{0, \pi\}$ and $\{\pi, \pi\}$. However,
the $\{0, \pi\}$ state, in the $J\rightarrow\infty$ case, consists of 
disconnected
chain-like structures. While the intra-chain arrangement is 
ferromagnetic, the inter-chain
arrangement is antiferromagnetic. Since there is no hopping 
connectivity between the
chains, the inter-chain exchange calculated in such a spin 
background
would emerge to be zero. Similarly, for a $\{\pi, \pi\}$ 
spin background, there are two sublattices
such that the intra-sublattice arrangement is ferromagnetic, 
while the inter-sublattice one
is anti-ferromagnetic. Again, since these sublattices are 
disconnected, the inter-sublattice
hopping is zero. In these anisotropic states the effective exchange
(and stiffness) vanishes along certain directions at $T=0$. 
To calculate the effective exchange that controls the $T_c$
in the AFM1 and AFM2 phases we need to necessarily solve the
finite temperature self-consistency problem.
This is an interesting problem, but computationally 
demanding, and is left for future work.

\subsection{$n-T$ phase diagram from the $D_{ij}$}

Our previous experience with the double exchange model 
suggests \cite{pm-scr} that a
reasonable estimate of  $T_{c}$ is provided 
by the exchange calculated in the fully FM $T=0$ state.
On this assumption, one can study the 
effective
spin model with classical Monte Carlo and calculate
finite temperature properties including $T_c$.
The  $n-T$ phase diagram obtained this way  is shown in 
the bottom panel of Fig~\ref{fig: Dij_Delta0_2}.
All the three phases: FM, 
AFM1, and AFM2
occur in approximately the correct filling windows. 
The $T_{c}$ scales for the ferromagnetic
phases, which are the only ones consistent with the assumed 
spin background, turn out
to be reasonably correct, as we will see in a comparison with the
full TCA result. We ignore the $T_c$ for the AF phase
since the AF exchange is not self-consistent.

\begin{figure}[t]
\includegraphics[width=5cm,height=6.5cm,angle=0,clip=true]{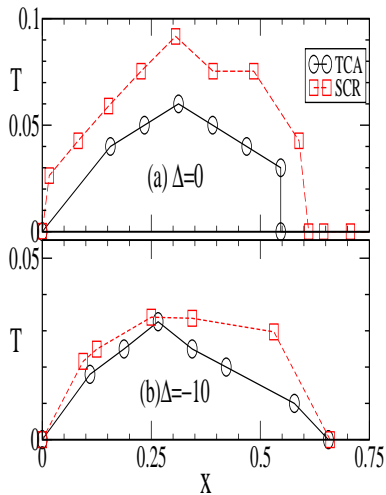}
\caption{Colour online:
Ferromagnetic part of the phase diagram for low `filling' \cite{footnote},
$x=3-n$, 
compared between TCA and
SCR, for (a) $\Delta=0$, (b) $\Delta=-10$ both calculated for 
size $16\times16$}  
\label{fig: phasediag_ferrocomp_TCA_SCR}
\end{figure}

\subsection{Properties in the ferromagnetic regime}

Since much of the interest in the double perovskites arises
from ferromagnetism, we focus on this regime in what 
follows.
In our $n-T$ phase diagram, this FM phase at low filling 
occurs upto 
 $n \approx 0.5-0.7$.  From the variational calculation, which is
essentially in the `bulk limit', the FM window is upto $\sim 0.3-0.4$.
Considering the degeneracy of the 
three $t_{2g}$ orbitals, translates
to about $0.9-1.2$ electrons per unit cell. Sr$_{2}$FeMoO$_{6}$, 
which has one electron per unit
cell, falls within this regime. However, many other materials 
like Sr$_{2}$FeReO$_{6}$ are
known, which have 2 or more electrons per unit cell, but are still 
ferromagnetic with a high
$T_{c}$. This discrepancy between theory and experiment was 
noticed by many authors before us:
Chattopadhyay and Millis~\cite{Millis}, 
L.Brey {\it et. al.}~\cite{SDSarma}, E. Carvajal {\it et.al.}~\cite{Avignon} 
and J.L.Alonso {\it et. al.}~\cite{FGuinea}. They attributed it to the
presence of competing antiferromagnetic channels, an effect 
which we also find. 

\begin{figure}[t]
\includegraphics[width=5cm,height=5cm,angle=0,clip=true]{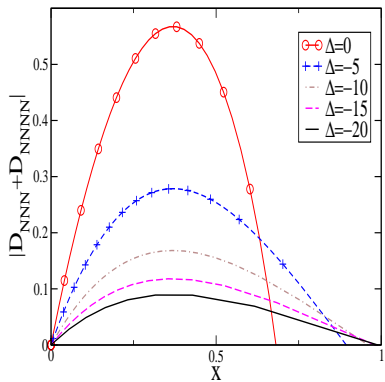}
\caption{Colour online:
Effective exchange in the ferro phase
$|D_{NNN}+D_{NNNN}|$ for different $\Delta$,  obtained from
SCR. The exchange calculation is on a $k$ grid 
$1000\times1000$, and the filling \cite{footnote} is $x= 3-n$.}
\label{fig: DNNN_vs_N_diff_Delta}
\end{figure}

Our results on  $T_{c}(n) $ is similar to 
that obtained by others, {\it i.e}, a reduction
as $n \rightarrow 0$ as the kinetic energy and ferromagnetic 
exchange weakens, and a drop also at large $n$ due to
the presence of competing AF phases. 
The $T_c(n)$ obtained from the SCR scheme is compared to
the result of full spin-fermion Monte Carlo using TCA,
Fig~\ref{fig: phasediag_ferrocomp_TCA_SCR}. They 
seem to match
quite well, except that the SCR results calculated on a 
$T=0$ state  overestimate  the
$T_c$ slightly. 
While the actual $T_{c}$-s can only be calculated using
Monte Carlo on small systems, eg. for $16 \times  16$ in 
Fig~\ref{fig: phasediag_ferrocomp_TCA_SCR},
the average exchange can 
calculated with a very large ${\bf k}$ grid 
($1000\times1000$ ${\bf k}$ points).
This $T=0$ exchange, based on an assumed FM state has the behaviour shown
\cite{footnote}  in Fig~\ref{fig: DNNN_vs_N_diff_Delta}. 
While this result has a clear 
correspondence with the $T_c(n, \Delta)$  
results obtained from DMFT  by  
Millis {\it et. al.}, and the calculations of Carvajal 
{\it et.al.}, it overestimates the window of FM, and misses 
the 
first order FM to AFM transition. The actual $T_c(x)$ will have a 
discontinuity with increasing $x$, instead of decreasing smoothly to zero.

\subsection{The AFM phases}

The AFM phases AFM1 and AFM2 occupy a large part of the 
$n-T$ phase diagram. The presence of such
collinear antiferromagnetic phases have been observed earlier 
by several authors, notably
Alonso {\it et.al.}~\cite{FGuinea}. 
These phases, at least within the $J\rightarrow\infty$
model considered here, have a a `lower 
connectedness' than the ferromagnetic phase. The AFM1 
phase consists of double staircase-like 
structures attached back to back, while
the AFM2 phase consists of decoupled 
Cu-O like lattices for each B'  spin channel.
The DOS corresponding to these are given in 
Fig~\ref{DOS_ferro_0pi_pipi}. The DOS for the
AFM1 phase resembles that of a 1D 
tightbinding lattice, while the AFM2 DOS is more 
2D-like. It is interesting to note that there is a 
dispersionless level for both the AFM phases,
which gives the jump in the $\mu-n$ curve. While the 
effective exchange calculation starting
from a fully polarized background already produced the 
three phases, but a truly self-consistent
calculation for the AFM phases would have to start 
assuming these spin backgrounds. Such a calculation is nontrivial,
as discussed before. 

\section{Magnetisation at the B' sites}

\begin{figure}[tbp]
\includegraphics[width=6cm,height=8cm,angle=0,clip=true]{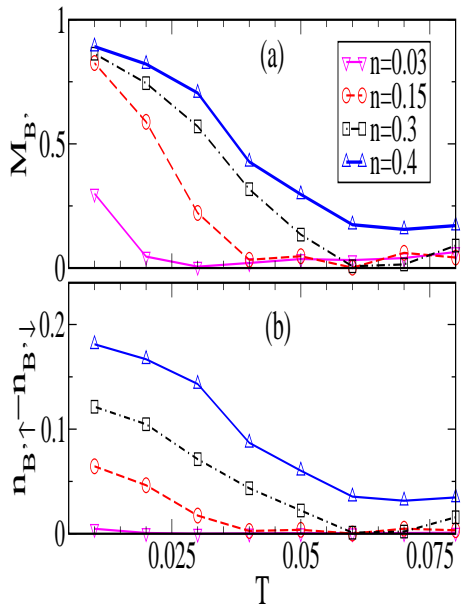}
\caption{Colour online:
Temperature dependence of B' site magnetization 
for different filling, 
from ED-MC on a $8 \times 8$ system at $\Delta=0$. 
The top panel plots the magnetisation normalized by the total B'
occupancy, while the bottom panel shows the unnormalized 
magnetisation.}
\label{fig: Momag_vs_T_diffN}
\end{figure}

\begin{figure}[tbp]
\includegraphics[width=6cm,height=5.5cm,angle=0,clip=true]{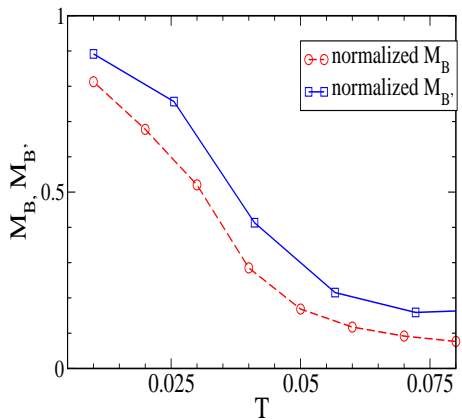}
\caption{Colour online:
Comparison of the electronic magnetisation at the B' site with
the core spin 
magnetization at the B site. The result is for  $n=0.4$ using  
ED-MC  on a  $8 \times 8$ system. Both the B and B' results are normalised 
to highlight their similar temperature dependence.}
\label{fig: compare_Mo_Fe_magN31}
\end{figure}

The induced magnetism on the B'  site was explained within 
a local `level repulsion' picture 
by Sarma {\it et al}~\cite{DDPRL}, but a lattice-oriented 
approach is lacking. Some headway
can be made by exactly integrating out the B degrees of 
freedom rather than the B' from the
$J\rightarrow\infty$ model given in Eq~\ref{Jinfinityham}, 
but the result is in the form of an
action, rather than an effective Hamiltonian. Within second 
order perturbation theory,
however, it appears that an extra onsite term of the
form
$\frac{zt_{FM}^{2}}{2\Delta}
\sum_{j\delta\alpha\beta} 
\vec{S}_{j+\delta}\cdot m_{j\alpha}^{\dagger}
\vec{\sigma_{\alpha\beta}}m_{j\beta}$
occurs for the 
B'  sites ($z$ is the number of nearest neighbours) giving 
 an `exchange splitting' at the
B' site, with the spins of the surrounding 
B sites serving as a magnetic field.
The corresponding hopping terms are given by:
\begin{eqnarray}
&&\frac{t^{2}_{FM}}{\Delta}\sum_{<ij>,\sigma}
(m_{i\sigma}^{\dagger}m_{j\sigma}
+\frac{t_{FM}^{2}}{\Delta}\sum_{<ij>}\vec{S}_{i+\vec{\delta}}\cdot 
m_{i\alpha}^{\dagger}\vec{\sigma}_{\alpha\beta}m_{j\beta}
\cr
&&
+\frac{t^{2}}{2\Delta}\sum_{<<ij>>}\vec{S}_{(i-j)/2}
\cdot m_{i\alpha}^{\dagger}\sigma_{\alpha\beta}m_{j\beta} 
\end{eqnarray}
Within a mean-field treatment of the B core spins 
${\bf S}_i \approx \langle {\bf S}_i \rangle 
= {\hat z} {M}$, the effective B'-B'  
`hamiltonian' can be written as:
\begin{eqnarray}
&&\frac{t^{2}}{\Delta}\sum_{<ij>}[(1+M)m_{i\uparrow}^{\dagger}
m_{j\uparrow}+(1-M)m_{i\downarrow}^{\dagger}m_{j\downarrow}]
\cr
&&+\frac{t^{2}}{2\Delta}\sum_{<<ij>>}[(1+M)
m_{i\uparrow}^{\dagger}m_{j\uparrow}+(1-M)
m_{i\downarrow}^{\dagger}m_{j\downarrow}]
\cr
&&
+\frac{2t^{2}}{\Delta}M
\sum_{i}(m_{i\uparrow}^{\dagger}m_{i\uparrow}-
m_{i\downarrow}^{\dagger}m_{i\downarrow})
\end{eqnarray}
Obviously, at $T=0$, $M=1$, and only one spin species hops.

The effective spin polarization at the B' 
site, which is purely electronic, contributes to the
total magnetization.  
Within the TCA approach using 
the Hamiltonian~\ref{Jinfinityham}, 
this can simply be estimated by calculating the 
normalized magnetization
$\frac{<n^{B'}_{\uparrow}>-<n^{B'}_{\downarrow}>}
{<n^{B'}_{\uparrow}>+<n^{B'}_{\downarrow}>}$.
 In Fig~\ref{fig: Momag_vs_T_diffN}(a), the 
magnetization of the B' has been plotted against 
the temperature for different fillings
correponding to an ED-MC simulation on a 8X8 
system for $\Delta=0$, normalized by the
net B' filling.
It is observed that the $T$ dependence 
is very similar to  that the
B  core spin case.
In Fig~\ref{fig: Momag_vs_T_diffN}(b), the bare 
B' magnetization is shown without
normalization: it shows that the saturation magnetization 
increases with filling, as expected. 
In Fig~\ref{fig: compare_Mo_Fe_magN31}, a comparison 
of the $M$ vs $T$ coming from the 
B' electron and the B core spins is provided.

\section{The effect of B'-B' hopping}

Inclusion of B'-B' hopping $t'$  would result in 
the same expression for the
exchange calculated in the ferromagnetic state as before 
(see Appendix A), except that
the $\Delta$ everywhere would get replaced by 
$\Delta+4t'cosk_xcosk_y$, while 
$\epsilon_{\bf k}=2t(cosk_x+cosk_y)$ for a square lattice. 
It is obvious that if $t=0$,
the exchange would vanish irrespective of $t'$, 
showing that hopping across the magnetic
site is crucial, as expected. For parameter values
reasonable in 
double perovskites,
$t' \approx0.1-0.3t$, and $0\le\Delta<3$, 
there is almost no change in the $n-\Delta$
phase diagram, although the $T_{c}$ values decrease marginally 
when the $t'$ is turned on.
For larger values of $\Delta$, the AFM1 phase becomes 
unstable, and the ferromagnetic window
extends a bit, upto the AFM2 phase, although 
the $T_{c}$s, of course, 
are proportionately low.

\section{Discussion}

We discuss a few issues below to connect our results to
available data on the double perovskites, and also
highlight a few  effects that we have neglected.

1.~{\it Material parameters:}
Ab initio calculations suggest \cite{DDPRL,Millis} that 
$t \sim 0.3-0.5$eV, while  
$t'$ is typically 3-5 times smaller.
The direct hopping between B sites is even smaller, $\sim 0.05$eV.
Estimates for the bare `charge transfer gap' 
$\Delta$ (in SFMO) vary between $1.4$eV~\cite{DDPRL}
to about $2$eV~\cite{Millis}. Hence, the parameter window
we explored seems 
reasonable. Our ferromagnetic $T_{c}$ are 
typically $0.1t$ at a filling appropriate to SFMO,
so the absolute magnitude of the $T_{c}$-s would be
about $360-600K$, roughly the range seen in the
for double perovskites.

2.~{\it B-B hopping:}
While we have not considered the effect of B-B hopping, the smallest 
energy scale in the problem, explicitly in this paper, it is 
possible to understand qualitatively the effect of including
this hopping. If only B-B'  hopping is considered, 
then there are two singular features in the density of states, 
at $\epsilon_{B}$ and $\epsilon_{B'}$, or alternatively, at $0$ and 
${\tilde \Delta}$. Inclusion of the B'-B' hopping 
resulted, at the zeroth level, in the smoothening of the feature 
at $\epsilon_{B'}$. Similarly, inclusion of the B-B hopping will 
basically smoothen out the feature at $\epsilon_{B}$. However, the  
B'-B' hopping
had a much more dramatic consequence in terms of providing an 
alternate pathway for delocalization of the B' 
electrons irrespective of spin, especially at large $\Delta$. Secondly,
it resulted in connecting up of the AFM1 staircases, 
getting rid of their 1D character,
and making this phase unstable compared to the FM and AFM2  
phases. Inclusion of a small B-B hopping in addition, 
is not, on the other hand, expected to have any more dramatic 
consequences. We can readily include this in our formalism.

3.~{\it Three dimensions:}
The entire analysis in this paper was in two dimensions. 
Apart from simplicity, ease of
visualization, and computational tractability, 
there is a definite argument in terms of the
symmetry of the $t_{2g}$ orbitals as long as nearest neighbour 
interactions are considered
~\cite{Millis,Avignon}, which says that one can consider three 
independent 2D Hamiltonians.  Other 
authors~\cite{ATaraphdar,Avignon2} have also used 2D Hamiltonians.
 Having said that, the phases discussed here generalizes easily
to three dimensions. The AFM1 phase in the 
absence of B'-B'  hopping 
becomes 2D rather than 1D, consisting of ferromagnetic [111] 
planes arranged antiferromagnetically. Such an arrangement has been 
observed by authors like
 Alonso {\it et. al.}~\cite{FGuinea}, and even 
in {\it ab initio} calculations
~\cite{abinitcondmat}.

4.~{\it Filling control:}
The $n-T$ phase diagram that we provide is  
for a definite set of parameters $\Delta,t,t'$ etc. 
While going across the series
of DP compounds Sr$_{2}$FeMoO$_{6}$, to Sr$_{2}$FeReO$_{6}$, it is 
not just the filling but
also all these parameters which are changing. A more controlled way of 
varying the filling alone
would probably be to dope the compounds at the A site, namely 
prepare the series Sr$_{2-x}$La$_{x}$FeMoO$_{6}$. 
While some work has been done in this
regard~\cite{Serrate}, more extensive work, 
probing higher doping values is necessary to 
ascertain whether such antiferromagnetic phases are indeed observed.

5.~{\it SCR for antisite disordered case:}
While the clean problem has been studied in detail in this paper, antisite disorder is expected
to make it more interesting. Paradoxical effects like increase in $T_{c}$ and widening
of the FM region has been suggested~\cite{FGuinea}. 
The scheme for self-consistent renormalization that we 
have proposed can be generalized even to the case of 
antisite disorder (see Appendix). However,
the scheme in that case becomes more 
numerical, and the analytical handle available here would
be lost even at $T=0$.
The formalism is presented in the Appendix and we are
currently studying the problem.

\section{Conclusion}

We have suggested a scheme for extracting a simple magnetic model
for double perovskites starting with a tight binding parametrisation of
the electronic structure. 
The `exchange scale' in this model is related to the electronic
kinetic energy. The ferromagnetic $T_c$ estimated from this
exchange compares well with results from the full spin-fermion
Monte Carlo.
The change in sign of the exchange with increasing carrier density
captures the phase competition in the electronic model and the 
transition in the magnetic ground.
Our scheme, extended to include multiple bands and spin-orbit coupling
would allow a controlled and economical approach to the finite
temperature physics of a wide variety of double perovskites.
Another fruitful line of exploration is to include antisite disorder.
We have highlighted the scheme in the appendix and hope to
present results in the near future.

\vspace{.3cm}

We acknowledge discussions with D. D. Sarma, Brijesh Kumar and Rajarshi 
Tiwari, and use of the Beowulf cluster at HRI.

\section{Appendix A: exchange calculation in the double perovskites}

The effective exchange $D_{ij}$
can be evaluated in the perfectly spin-ordered state at T=0.
This can be obtained by using the known form of the Green's function 
$\left<\overline{\gamma}_{in}\gamma_{jn}\right>$ at T=0, namely it is the
$G_{ff0}(\vec{r}_{i}-\vec{r}_{j},i\omega_{n})$ obtained before, 
made dimensionless by dividing
by $\beta$. Hence, in the spin ordered case, the exchange becomes:
$$U_{B}(T=0)=\frac{1}{\beta}\sum_{n}
\frac{(2i\omega_{n}-\Delta)}{(i\omega_{n}-\Delta)^{2}}
\frac{\epsilon_{k}^{2}}{\left[i\omega_{n}-
\frac{\epsilon_{k}^{2}}{i\omega_{n}-\Delta}\right]} \nonumber $$ 
\begin{eqnarray}
=\frac{1}{\beta}\sum_{n}\frac{(2i\omega_{n}-\Delta)
\epsilon_{k}^{2}}{(i\omega_{n}-\Delta)
[i\omega_{n}(i\omega_{n}-\Delta)-\epsilon_{k}^{2}]}
\label{Dij_T0_iwn}
\end{eqnarray}

Expanding in partial fractions, this can be written as:
\begin{equation}
U_{B}=\frac{1}{\beta}\sum_{kn}\left[\frac{E_{k+}}
{i\omega_{n}-E_{k+}}+\frac{E_{k+}}{
i\omega_{n}-E_{k-}}-\frac{\Delta}{i\omega_{n}-\Delta}\right]
\label{U_B_T0_iwn}
\end{equation}
Performing the Matsubara sums, this gives the final result
\begin{equation}
U_{B}=\sum_{k}\left[E_{k+}n_{F}(E_{k+})+E_{k-}n_{F}(E_{k-})\right]
-\Delta n_{F}(\Delta)
\label{Dij_T0_k}
\end{equation}

The last term is an additional contribution obtained from the 
missing $B^{'}$ energy.
While this gives the full internal energy in the spin polarised case,
the bond-resolved exchange $D_{ij}$ is given by
\begin{eqnarray}
D_{ij} &=& \sum_{k}e^{i\vec{k}\cdot(\vec{r}_{i}-\vec{r}_{j})}
\left[E_{k+}n_{F}(E_{k+})+E_{k-}n_{F}(E_{k-})\right] \cr
&&~~~~~~~~~~~~~~~~~~~~-\Delta n_{F}(\Delta)
\end{eqnarray}

We need to do the k-sum, which
can only be done numerically for square (2D) or cubic (3D) lattices. 
Instead, if one uses
a Bethe lattice of infinite coordination, then the bare density of 
states for nearest neighbour
hopping is semicircular, and analytic 
treatment is possible, at least for the DOS. 
In our case, there are two bands,
with dispersions 
\begin{equation}
E_{k\pm}=\frac{\Delta\pm\sqrt{\Delta^{2}+4\epsilon_{k}^{2}}}{2}
\label{ak_pm_tFM_only}
\end{equation}
The density of states for these bands is:
\begin{eqnarray}
\rho_{\pm}(E) &=&\sum_{k}\delta
\left(E-\frac{\Delta\pm\sqrt{\Delta^{2}+4\epsilon_{k}^{2}}}{2}
\right) \cr
 &=&\int d\omega\rho_{0}(\omega)\delta\left(E-\frac{\Delta\pm\sqrt{\Delta^{2}+
4\omega^{2}}}{2}\right)
\label{rho_ak_pm}
\end{eqnarray}
where the bare DOS $\rho_{0}(\omega)=\sum_{k}\delta(\omega-\epsilon_{k})$ 
is approximated by the 
semicircular DOS, $\rho_{0}(\omega)
\approx\frac{2}{\pi D^{2}}\sqrt{D^{2}-\omega^{2}}$.
Then, the full DOS given by Eq~\ref{rho_ak_pm} becomes:
\begin{equation}
\rho_{\pm}(E)=\frac{2}{\pi D^{2}}\int_{-D}^{D}
d\omega\sqrt{D^{2}-\omega^{2}}\delta\left(E-\frac{\Delta\pm\sqrt{\Delta^{2}+
4\omega^{2}}}{2}\right)
\end{equation}
This integral can be evaluated, to give the 
following analytic result for the DOS:
$$\rho_{\pm}(E)=\frac{\pm(2E-\Delta)}{\pi D^{2}}
\frac{\sqrt{D^{2}-E^{2}+E\Delta}}{\sqrt{E^{2}-E\Delta}}$$
\begin{equation}
\times\Theta(D^{2}-E^{2}-E\Delta)\Theta(E^{2}-E\Delta)
\end{equation}
 Obviously, this diverges at $E=0$, 
and $E=\Delta$. The limits of the DOS, 
and hence the integral, are obtained from the
 equations given by the two theta function conditions:
$ 
E_{limit}^{(1))}=\frac{\Delta\pm\sqrt{\Delta^{2}+4D^{2}}}{2}$,
$ E_{limit}^{(2))}=0,\Delta $
If we take $\Delta<0$, as in our case, 
then the lower band lies between 
$\frac{\Delta+\sqrt{\Delta^{2}+4D^{2}}}{2}$ 
(left edge) and $\Delta$ (right edge), while the upper band lies between
$0$ (left edge) and $\frac{\Delta-\sqrt{\Delta^{2}+4D^{2}}}{2}$ (right edge).
 
 When $D\rightarrow0$, i.e., $t \rightarrow0$,
 the theta function conditions are only satisfied together
for $E^{2}-E\Delta=0$, i.e., at $E=0$ or $E=\Delta$.
Thus we recover the bare levels as 
$\delta$-function peaks in the DOS, as expected.

The exchange at T=0 is given in terms of this DOS as:
\begin{equation}
U_{B}=\int E[\rho_{+}(E)+\rho_{-}(E)]n_{F}(E)dE-\Delta n_{F}(\Delta)
\label{U_rho}
\end{equation}
This gives $U_{B}$ as a function of $\mu$. 
One can also obtain the total number of
electrons $N$ from the DOS as a function of $\mu$:
\begin{equation}
N=\int dE[\rho_{+}(E)+\rho_{-}(E)]n_{F}(E)
\label{N_rho}
\end{equation} 
From Eq~\ref{U_rho} and Eq~\ref{N_rho}, 
eliminating $\mu$, one can get $U$ vs $N$.

Using the substitution $\omega=E^{2}-E\Delta$, 
the expression for the exchange can be
rewritten in a more convenient form at T=0 as:
\begin{eqnarray}
U_{B}(\mu)&=&
\frac{2}{\pi D^{2}}\sum_{\pm}\int_{-D}^{D}d\omega
\frac{\Delta\pm\sqrt{\Delta^{2}+4\omega^{2}}}{2} \cr
&& \times\Theta\left(\mu-\frac{\Delta\pm
\sqrt{\Delta^{2}+4\omega^{2}}}{2}\right)\sqrt{D^{2}-\omega^{2}} \cr
&& ~~~~-\Delta\Theta(\mu-\Delta)
\label{U_B_subs}
\end{eqnarray}

It is to be noticed that as $t \rightarrow0$, 
i.e., $D\rightarrow0$, the exchange goes
to zero, as it should. 
This can be seen in two ways. Firstly, as $t \rightarrow0$,
$\epsilon_{k}\rightarrow0$. Hence, the band dispersions $E_{k\pm}$
given by Eq~\ref{ak_pm_tFM_only} tends to $\frac{\Delta\pm|\Delta|}{2}$. 
This means that $E_{k+}\rightarrow0$ and 
$E_{k-}\rightarrow\Delta$. Hence, putting in Eq~\ref{U_rho},
 the first term involving $E_{k+}$ is $0$, 
while the second term involving $E_{k-}$
 cancels the third term involving $\Delta n_{F}(\Delta)$.

The other way to observe this is to use 
Eq~\ref{U_B_subs}. Here, when $D\rightarrow0$,
then the theta function condition is 
only satisfied for $\omega^{2}=D^{2}$, which means that
 the only contribution to the integral comes 
from $\omega=0$. Indeed, the bare semicircular
 DOS $\frac{2}{\pi D^{2}}\sqrt{D^{2}-\omega^{2}}$ 
being a normalized object, tends to a
 delta function $\delta(E)$ as $D\rightarrow0$. 
Hence, the term involving $+$ sign gives $0$,
 while that involving $-$ sign gives $\Delta\Theta(\mu-\Delta)$, 
 which cancels with the third term.

\section{Appendix B: exchange calculation with 
antisite disorder}

Firstly, the Hamiltonian has to be written in such a way that all the B and B' 
degrees of freedom are separated out in distinct subspaces of the Hamiltonian.
$$
H=\left(\begin{array}{cc}
H_{FF} & H_{FM} \\
H_{MF} & H_{MM} \\
\end{array}\right)
$$

where $H_{FF}$ represents the terms in the subspace of the B 
degrees of freedom, while $H_{MM}$ represents the terms in the 
B' subspace. $H_{MF}$ and $H_{FM}$ connects
the two subspaces.

The B Green's function satisfies the matrix equation
\begin{equation}
G_{FF}^{-1}(i\omega_{n})=
i\omega_{n}\mathbf{I}-H_{FF}-H_{FM}(i\omega_{n}\mathbf{I}-H_{MM})^{-1}H_{MF}
\end{equation}

Written out term by term,

\begin{equation}
G_{FF}^{-1}(i\omega_{n})=
i\omega_{n}\delta_{ij}-H_{FF_{ij}}-\sum_{kl}H_{FM_{ik}}
(i\omega_{n}\mathbf{I}-H_{MM})^{-1}_{kl}H_{MF_{ij}}
\end{equation}

Hence the action 
\begin{eqnarray}
&& {\mathcal A}\{S\}=
\sum_{in}(i\nu_{n}+\beta\mu)\overline{\gamma}_{in}\gamma_{in}-
\sum_{ij}\beta H_{FF_{ij}}\overline{\gamma}_{in}\gamma_{jn}g_{ij}
\cr
&&
%\right
-\beta^{2}\sum_{kl}H_{FM_{ik}}(i\nu_{n}
\mathbf{I}-\beta H_{MM})^{-1}_{kl}H_{MF_{ij}}
%\right]
\overline{\gamma}_{in}\gamma_{jn}g_{ij}
\nonumber
\end{eqnarray}

Now, let $V$ be the diagonalizing matrix of $H_{MM}$ 
and its eigenvalues are $\lambda^{p}_{MM}$.
Then,
\begin{eqnarray}
\mathcal{A}\{S\} &=&
\sum_{in}(i\nu_{n}+\beta\mu)\overline{\gamma}_{in}\gamma_{in}
-\sum_{ijn}\left[\beta H_{FF_{ij}}\right. \cr
&& 
\left.-\beta^{2}\sum_{klp}\frac{H_{FM_{ik}}V_{kp}V_{pl}^{-1}H_{MF_{lj}}}
{i\nu_{n}-\beta\lambda^{p}_{MM}}\right]g_{ij}\overline{\gamma}_{in}\gamma_{jn}
\nonumber
\end{eqnarray}

\begin{eqnarray}
&&\left<\frac{\partial A}{\partial \beta}\right>=
\sum_{ij}\left[H_{FF_{ij}}-
\sum_{klp}\frac{H_{FM_{ik}}V_{kp}V_{pl}^{-1}H_{MF_{lj}}}
{i\nu_{n}-\beta\lambda^{p}_{MM}}+\right.
\cr
&&
\left.\beta^{2}\sum_{klp}\frac{H_{FM_{ik}}V_{kp}
\lambda^{p}_{MM}V_{pl}^{-1}H_{MF_{lj}}}
{(i\nu_{n}-\beta\lambda^{p}_{MM})^{2}}\right]
g_{ij}<\overline{\gamma}_{in}\gamma_{jn}>
\cr
&&=-\sum_{ij}D_{ij}g_{ij} 
\nonumber
\end{eqnarray}

It is to be noted that in the ordered case, the hamiltonian 
matrix becomes block diagonal in
k-space, and the elements of the off-diagonal 
block $H_{MF}$ and $H_{FM}$ are simply 
$\epsilon_{k}$, while those of the digonalizing matrices 
$V$ are $e^{i\vec{k}.(\vec{r}_{i}-\vec{r}_{j})}$, 
while the eigenvalues $\lambda^{p}_{MM}$ are simply $\epsilon_{Mo}$, 
i.e., $\Delta$.
Hence, the quantity $H_{FM_{ik}}V_{kp}V_{pl}^{-1}H_{MF_{lj}}$ 
in the numerator 
simply goes over to $h_{ij}$, as defined in Eq~\ref{spinmodel}


\begin{thebibliography}{9}
\bibitem{footnote} The x-axis  
represents $(3-n)$ rather
than $n$. This is done to avoid numerical problems associated 
with the calculation of the
exchange on the 
low filling side due to the singularity at $\mu=\Delta$. 
For small $\Delta$, of course,
there is an explicit symmetry between the two sides, but for 
larger $\Delta$, the ferromagnetic
phase seems to have a larger extent on the high filling side 
compared to the low filling one.
\end{thebibliography}
\end{document}